\documentclass{ws-p8-50x6-00}
\usepackage{amsmath}

\def\bb{{b\bar{b}}}
\def\cc{{c\bar{c}}}
\def\ll{{l\bar{l}}}

\def\PL{Phys.\ Lett.\ {\bf B}}

\def\PRL{Phys.\ Rev.\ Lett.\ }

\def\NIM{Nucl.\ Instr.\ Methods\ }

\def\Abreu{DELPHI Coll., P. Abreu {et al.,}\ }

\begin{document}

\title{TWO TOPICS IN MULTIPARTICLE DYNAMICS AT LEP: MULTIPLICITY IN
$\lowercase{b\bar{b}}$ EVENTS AND\\ 
SCREWINESS AT THE END OF THE QCD CASCADE}

\author{ALESSANDRO DE ANGELIS}

\address{CERN, Geveva, Switzerland}

\maketitle

\vspace*{-7.5cm}
\begin{tabular*}{15.cm}{l@{\extracolsep{\fill}}r}
  & 
CERN-OPEN 98-031
\\
& 
19 November, 1998
\\
&\\
\end{tabular*}
\vspace*{6.cm}

\abstracts{
This talk deals with two topics in multiparticle dynamics investigated by means of the DELPHI detector at LEP.
Related to the first topic,
we have used the data  collected at 183 GeV to 
measure the average charged particle multiplicity 
in $e^+e^- \rightarrow b\bar{b}$ events.
The result is remarkably in agreement with QCD predictions, while it is 
more than two standard deviations larger than calculations assuming that 
the multiplicity accompanying the decay of a heavy quark is independent 
of the mass of the quark itself. 
The second topic deals with a recent theoretical model by 
Andersson et al., in which  
soft gluons order themselves in the form of a helix
at the end of the QCD cascades. 
In our data at the Z peak, we have found no evidence for such an effect.}

\section{Charged Particle Multiplicity in $b\bar{b}$ Events}

The study of the properties of 
the fragmentation of heavy quarks compared to
light quarks  offers insights in perturbative QCD. Particularly important is
the difference in charged particle multiplicity between 
light quark and heavy quark initiated events in
$e^+e^-$ annihilations. QCD predicts
(see \cite{schumm,DOKBM,petrov,deus} and \cite{khoze} for a recent review),
somehow counter-intuitively \cite{kisselev},
that this difference is energy independent; this is motivated by
mass effects on the hadronization.

The present experimental tests, although preferring the QCD-motivated
scenario, were not conclusive 
(see \cite{schumm} and references therein,
\cite{delphi,opal,sld,tristan}). 
At LEP 2 energies, however, the difference 
between the QCD prediction and the model ignoring mass effects 
is large, and the experimental measurement can firmly distinguish between the
two hypotheses.

Data corresponding to
a total luminosity of 54 pb$^{-1}$ collected by DELPHI 
at centre-of-mass energies around 183~GeV during 1997 were analysed.
A description of the DELPHI detector can be found in \cite{deldet}; its
performance is discussed in \cite{perfo}.

Hadronic events were selected as in \cite{noi}.
The cross-section for $e^+e^- \rightarrow {\mathrm q}\bar{\mathrm q}(\gamma)$ 
above
the Z peak is dominated by radiative q$\bar{\mathrm q}\gamma$ events,
such that the effective hadronic
centre-of-mass energy, $\sqrt{s'}$, is generally
smaller than $\sqrt{s}$. Only events with $\sqrt{s'}$ (determined using the
procedure described in \cite{sprime}) larger than 170 GeV were used.

The influence of the detector on the analysis was studied with
the full DELPHI simulation program, DELSIM~\cite{perfo}.
Events were generated with
PYTHIA 5.7 and JETSET 7.4~\cite{lund} Parton Shower (PS).
The particles were followed through
the detailed geometry of DELPHI giving simulated digitisations in each
subdetector. These data were processed with the same
reconstruction and analysis programs as the real data.

A 90\% $b$ enriched sample has then been obtained by requiring that the
$b$-event tagging variable $y$ defined as in Ref. \cite{perfo} was lower 
than $5 \times 10^{-5}$. 
From this sample the average $\bb$ multiplicity has been measured by
unfolding via simulation for detector effects, selection criteria 
and initial state radiation. 
The value obtained was $<n_\bb> = 29.54 \pm 0.79 (stat)$.

The relative systematic uncertainty on $<n_\bb>$ was assumed as in 
\cite{delphi}, so that:
\begin{displaymath}
<n_\bb> = 29.54 \pm 0.79 (stat) \pm 0.63 (syst) \, .
\end{displaymath}
It was verified that the value above is stable within $\pm$ 0.13 units with
respect to variations of the cut on the $y$ variable.

The difference $\delta_{bl}$ between the $\bb$ multiplicity 
and the multiplicity in generic light quark $l = u,d,s$ events 
can then be extracted by using the equation:
\begin{displaymath}
<n_{ch}> =  R_b <n_\bb> +  R_c <n_\cc> +  R_l <n_\ll>  \, ,
\end{displaymath}
where the fractions $R_b=0.16$, $R_c=0.26$, $R_l=0.58$ are the ones 
predicted by the Standard Model.
For the $\cc$ multiplicity, the PYTHIA prediction of 27.6 was 
assumed, with a systematic error of $\pm 1.5$ units.
For $<n_{ch}>$, we used the result published by DELPHI in \cite{noi}:
\begin{equation}\label{mul183}
<n_{ch}> = 26.58 \pm 0.24 (stat) \pm 0.54 (syst)  \, .
\end{equation}

Finally we obtain:
\begin{equation}\label{delbl}
\delta_{bl} = 4.23 \pm 1.09 (stat) \pm 1.40 (syst) \, .
\end{equation}
The value assumed for $<n_\cc>$ is consistent within better than 1.5 
standard deviations with both $<n_\ll>$ and $<n_\bb>$,
neglecting the errors on $<n_\ll>$, $<n_\bb>$: we thus 
verified a-posteriori that the systematic error assumed for 
$<n_\cc>$ was conservative.
The dependence of $\delta_{bl}$ on $<n_\cc>$ is 
anyhow small, and the systematic error assumed 
gives a contribution of $\pm 0.67$ units 
to the systematic error on $\delta_{bl}$.


{\bf{Flavour-Independent Fragmentation ---}} 
In a model in which the hadronization is independent of the mass,
one can assume that the non-leading multiplicity
in an event,
i.e., the light quark multiplicity which accompanies 
the decay products of the primary hadrons,
is governed by the effective energy available
to the fragmentation system following the
production of the primary hadrons \cite{kisselev}.  
One can thus write:
\small{\begin{eqnarray*}
 \delta_{bl}(E_{cm}) & = & 2<n^{(decay)}_B>  + 
 \int_0^1  dx_B f(x_B) \int_0^1 dx_{\bar{B}} f(x_{\bar{B}}) \, \,
 n_{l\bar{l}}\left( \left(1-\frac{x_B+x_{\bar{B}}}{2} \right) 
 E_{cm}\right)\nonumber \\
 & - & n_{l\bar{l}}(E_{cm}) \, ,
\end{eqnarray*}}
where 
$<n^{(decay)}_B>$ is the average number of charged 
particles coming from the
decay of a $B$ hadron,
$x_B$ ($x_{\bar{B}}$) is the fraction of the beam energy taken by
the $B$ ($\bar{B}$) hadron, and $f(x_B)$ is the $b$ fragmentation function.

We assumed $2<n^{(decay)}_B> = 11.0 \pm 0.2$ \cite{schumm}, 
consistent with the average $<n^{(decay)}_B> = 5.7 \pm 0.3$
measured at LEP \cite{vietri}.
For $f(x_B)$, we assumed a Peterson function of average $0.70 \pm 0.02$
\cite{pdg}.
The value of $n_{l\bar{l}}(E)$ was computed from the fit
to a QCD formula \cite{webber} including the resummation of leading
(LLA) and next-to-leading (NLLA) corrections,
which reproduces well the 
measured charged multiplicities \cite{noi}, with appropriate 
corrections to remove the effect of heavy quarks \cite{dea}.

\begin{figure}
\mbox{\epsfxsize10cm\epsffile{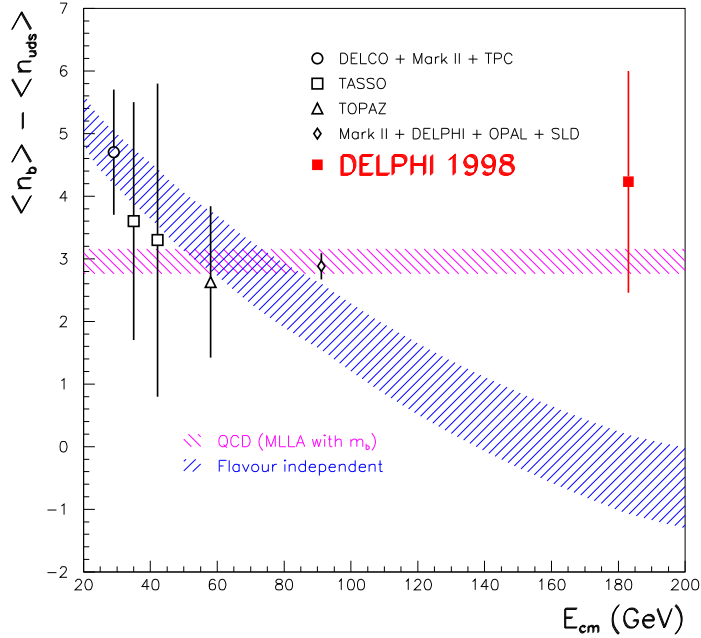}}
\caption[]{
The present measurement of $\delta_{bl}$ 
compared to previous measurements as a function of the centre-of-mass energy,
to the QCD prediction, and to the prediction from 
flavour-independent 
fragmentation.}\label{nice}
\end{figure}

The prediction of the model in which the hadronization
is independent of the mass is plotted in Figure \ref{nice}
as a function of the centre-of-mass energy.
There are several variations of the
model in the literature, 
leading to slightly different predictions
(see \cite{vietri} and references therein).

\noindent{\bf{QCD Calculation ---}} 
The large mass of the $b$ quark,
in comparison to the scale of the strong interaction,
$\Lambda \approx 0.2$ GeV, results in a natural 
cut off for the emission of gluon bremsstrahlung,
thereby avoiding the treatment of divergences
in perturbation theory.
Furthermore, where the centre-of-mass energy 
greatly exceeds the scale of the $b$ quark mass, 
the inclusive spectrum of heavy quark production is
expected to be well described by perturbative QCD
in the Modified Leading Logarithmic Approximation (MLLA).

The value of $\delta_{bl}$
has been calculated in perturbative QCD\cite{schumm,petrov}:
$$\delta_{bl} =  
2<n^{(decay)}_B>
- <n_{l\bar{l}}(e^{1/2}m_b)> +  O(\alpha_s(m_b))<n_{l\bar{l}}(m_b))> \, .$$
Although the last term in the equation limits the accuracy in
the calculation of $\delta_{bl}$, one can conclude that
$\delta_{bl}$ is approximately 
independent of $E_{cm}$. 

The calculation of the actual value of
$\delta_{bl}$ in \cite{schumm}
gives
a value of $5.5 \pm 0.8$; an approximation including the last term in the
above equation gives a value of 3.68 \cite{petrov}.
This demonstrates the importance of the contribution
$O(\alpha_s(m_b))$.
A condition less restrictive is the calculation of upper limits.
An upper limit $\delta_{bl} < 4.1$ is given in \cite{petrov}, while
from phenomenological arguments, 
$\delta_{bl} < 4$ is predicted in Ref. \cite{deus}. 

In Figure \ref{nice} the high energy prediction from QCD is taken from the
average of the experimental values of $\delta_{bl}$ up to $m_Z$
included, $<\delta_{bl}> = 2.96 \pm 0.20$.

Our  measurement of $\delta_{bl}$ is fully consistent with the
hypothesis of energy independence, and
more than two standard deviations larger than
predicted by the naive model presented in the beginning of this
section.

\section{Is There Screwiness at the End of the QCD Cascades?}
Two colour charges moving apart from each other produce colour radiation
in the form of 
gluons. At the end of the parton
cascade one enters into a non-perturbative region,
where perturbative QCD breaks up due to the small momentum transfer and hence large strong
coupling constant $\alpha_s$. 
There are non-perturbative models to describe the transition regime 
between the perturbative region and final state hadrons, in
which the largest contributing Feynman diagrams are chosen based on the coherence of the gluons.
The gluon emissions are ordered by the transverse momentum in the
dipole model~\cite{dipole}, and by emission angle in the Webber-Marchesini 
model~\cite{wm}. The azimuthal angle of the gluons is taken to be uniformly distributed
ignoring possible spin effects.

In a recent theoretical paper~\cite{skrewpap}, it is studied what happens at the 
end of the QCD cascades: the soft gluons have a transverse momentum
of the same order as the mass of the emitter, 
leading to a situation in which a large
recoil may spoil the original coherence. 
With the assumptions that: 
\begin{itemize}
\item the effective coupling  is large enough so that the number of
emitted gluons is as large as possible; 
\item the emissions conserve helicity;
\end{itemize}
the Authors show that the closest 
packing of soft gluons is obtained
by using the extra degree of freedom 
provided by the azimuthal angle and arranging
the gluons along a helix.
The helix is described by the parameter $\tau=dy/d\phi$, where 
$y$ is rapidity and $\phi$
is the azimuthal angle with respect to the original parton direction. The correlation
between the azimuthal angle and rapidity can be tested by using a Fourier power spectrum,
which is called screwiness $S$ in Ref.~\cite{skrewpap} and defined as:
\begin{equation}
S(\omega) = \sum_e P_e |\sum_j \exp(i(\omega y_j - \phi_j))|^2,
\end{equation}
where the second sum goes over the gluons, the first sum goes over all the events, and
$P_e$ is an event weight.
Screwiness would then manifest itself as a peak in the power spectrum at $\omega = 1/<\tau>$.
If a large number of gluons were emitted isotropically, 
there should be no peak. Asymptotically screwiness should only
depend of the number of emitted gluons at large values of $\omega$.

Experimentally one observes stable final state particles instead of gluons.
Moreover, in such an analysis it is convenient to use charged particles only,
due to the better accuracy in the measurement of the momenta at low energy. 
Normalized screwiness can now be defined as
\begin{equation}
S(\omega) = \frac{1}{N_{event} <N_{particle}>}\sum_e |\sum_j \exp(i(\omega y_j - \phi_j))|^2,
\end{equation}
where the second sum goes over the final state particles, 
and the first sum goes over all the events. Rapidity 
$y$ and azimuthal angle $\phi$ are defined with respect to the thrust axis.
Ordering of the final state particles would produce a peak in the $S(\omega)$
spectrum, with a maximum peak size of $S(\omega)_{max} =<N_{particle}>$ 
for maximal ordering. An isotropical 
particle emission would give asymptotically 
$S(\omega) \simeq 1$ at large values of $\omega$, and no peak. 
When $\omega$ approaches zero, screwiness measures the 
transverse momentum balance of the final state particles and it should approach zero.

For a fixed $\sqrt{s}$, a certain finite rapidity interval is available.
The central rapidity plateau is not flat, producing
fluctuations in the power spectrum. Selecting the central rapidity region
and using only charged particles with experimental acceptance cuts,
the number of final state particles available to calculate the screwiness is reduced.
\begin{figure}
\begin{center}
\mbox{\epsfig{figure=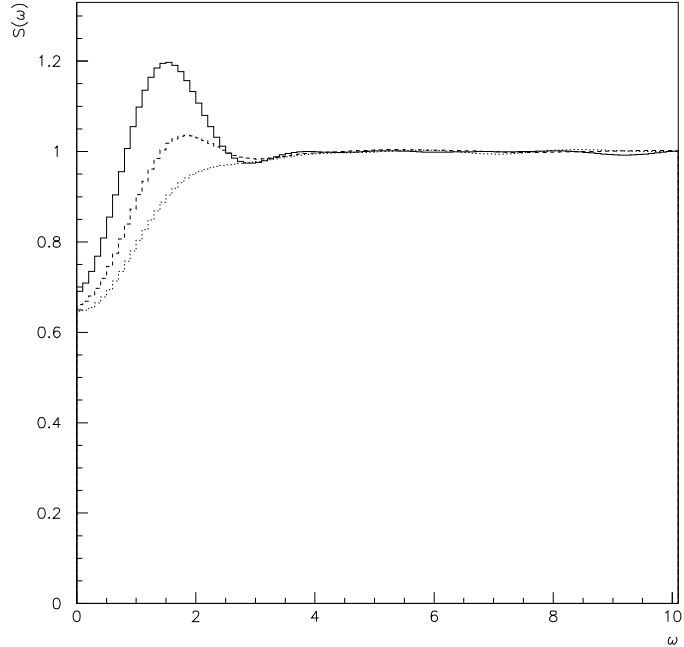,width=10cm}}
\protect\caption{\label{fig2} Power spectrum 
for simulated data samples with $\tau = 0.7$ (full line), $\tau = 0.5$ (dashed line) 
and $\tau = 0.3$ (dotted line). Only stable charged 
particles with $p>0.15$~GeV/$c$ in $|y|<2$ were retained.}
\end{center}
\end{figure}

A typical screwiness spectrum is shown in 
Figure~\ref{fig2} where the parameter $\tau$ was set to 0.3, 0.5 and 0.7.
Only charged particles with $p>0.15$~GeV/$c$ in $|y|<2$ were retained.
A clear peak can be observed at $\omega = 1/<\tau>$ when $\tau = 0.7$. With a $\tau$ value of
0.5 there is still peak in the spectrum which could be observable, while with $\tau = 0.3$
the winding is too fast to be observed.
The events were generated using 
a JETSET Monte Carlo simulation program, which did not
contain parton showering, but in which a helix-ordering of the final state clusters was
implemented.
The average multiplicity of charged particles with the above 
selections was 5.2, which was roughly the same as with the default JETSET.

With soft gluons included in the event generator, 
the transverse momentum fluctuations event-by-event cause that the screwiness does not decrease
when $\omega$ goes to zero if one considers a narrow window in rapidity space with a few particles.
The multiplicity of particles in the event affect the 
behaviour of the screwiness at small values of $\omega$. At large values of $\omega$ 
the normalized screwiness approaches  one.
\begin{figure}
\begin{center}
\mbox{\epsfig{figure=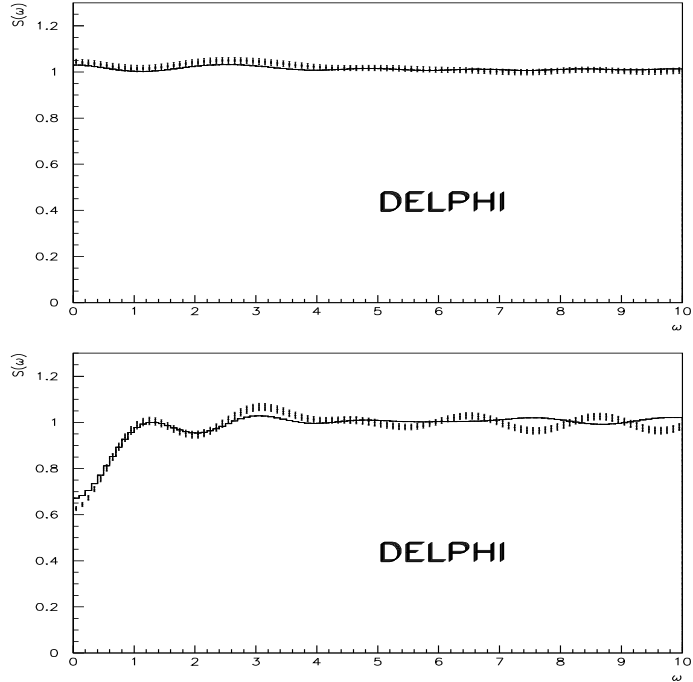,height=10cm,width=10cm}}
\protect\caption{\label{fig3} Upper: power spectrum  
for DELPHI data (points) 
compared to JETSET with full detector simulation (histogram).
The inner error bars on the data points correspond to the statistical
errors, the outer ones to the sum in quadrature of the statistical
and systematic errors. Lower: 
as for the upper part, for events with at least five charged particles and 
a missing momentum 
component transverse to the thrust axis smaller than 1~GeV/$c$.
}
\end{center}
\end{figure}

A sample of hadronic Z decays collected in 1994
were selected by requiring that the events contained
at least five charged particles within the angular 
range of $|\cos\theta|<0.96$,  
that the thrust axis was within the angular range of $|\cos\theta_{T}|<0.7$,
and that the thrust value was larger than 0.98.
Charged particles were accepted if
the measured momentum was greater than 0.15~GeV/$c$, 
track length greater than 30 cm and the
transverse and longitudinal impact parameters 
less than 5 and 10 cm, respectively. Particles within $|y|<2$ 
were retained for the calculation of the power spectrum.
The final selected sample used
for this analysis consisted of 50,000 events.

Screwiness of the selected events is shown in the upper part of 
Figure~\ref{fig3} together with the full simulation. 
The simulated data were generated with the JETSET 
7.3 parton shower simulation program, using input parameters reproducing the main features of the
DELPHI data~\cite{param}. The generated events were then 
simulated with the DELSIM full detector simulation program~\cite{perfo} and
reconstructed as data. The data show no peak in the
screwiness distribution, and the JETSET model, with random
emission of soft gluons in the azimuthal angle, 
reproduces well the data distribution.

The statistical error of the data was about 0.5\%, 
computed as the standard deviation of a set
of Monte Carlo simulation samples with the same population as the data. 
The systematic uncertainty was estimated to be around 1\%. 
The main source of systematic uncertainty was
expected to originate from the small differences in the rapidity distribution in
data and simulation. The following procedure was used to
estimate the inability of the simulated events to reproduce the 
data. The data were first weighted in
such a way that the rapidity distribution was flat. 
The data were then weighted with the weights 
which would make the simulated rapidity distribution flat. The difference 
of the results, divided by the data distribution, was then the
estimate of the relative systematic error.

The same analysis was repeated by selecting only events with at least five 
charged particles
with $|y| < 2$, but no effect was seen. 
Finally, 10,000 events were selected with at least five charged particles
with $|y| < 2$ and such that the sum of the momenta of these particles had
a momentum component transverse to the thrust axis smaller than 1~GeV/$c$ 
(lower part of Figure~\ref{fig3}). The disagreement between data and 
simulation after these cuts
is at the 5\% level. The data does not show, however, 
a typical screwiness pattern.



In conclusion, no evidence for screwiness was found, 
and results were consistent with random emission
of gluons in the azimuthal angle.
Given the DELPHI experimental sensitivity,
it was found that the helix parameter $\tau=dy/d\phi$ cannot be larger than about 0.5.

\subsection*{Acknowledgements}
I thank the organizers for the pleasant time in Greece.

I am grateful to V. Khoze and to V. Petrov for
useful discussions, and to F. Cossutti for proofreading. 

The material presented in the first Section of this talk comes from a work
with P. Abreu, R. Nobrega, M. Pimenta and L. Vitale \cite{bnote};
the material presented in the second Section comes from a work
with P. Eerola and M. Ringn\'er \cite{snote}. Special thanks to them.


\end{document}